\documentclass[a4paper]{PoS}

\usepackage{amsmath}
\usepackage{amssymb}
\usepackage[utf8]{inputenc}
\usepackage{xspace}
\usepackage[american]{babel}
\usepackage{times}
\usepackage{graphicx}
\usepackage{multirow}
\usepackage[numbers]{natbib}
\usepackage{rotating}

\newcommand{\ie}{i.\,e.\xspace}

\newcommand{\eg}{e.\,g.\xspace}
 
\newcommand{\GRB}{Gamma-Ray Burst\xspace}
\newcommand{\GRBs}{Gamma-Ray Bursts\xspace}
\newcommand{\ept}{\epsilon(\theta)}
\newcommand{\tv}{\theta_\mathrm{v}}
\newcommand{\tc}{\theta_\mathrm{c}}
\newcommand{\tj}{\theta_\mathrm{jet}}
\newcommand{\Eiso}{E_\mathrm{iso}}
\newcommand{\Liso}{L_\mathrm{iso}}

\newcommand{\dex}{\thinspace\mathrm{dex}}

\newcommand{\tgtc}{\theta_\Gamma/\tc}

\def\gsim{ \lower .75ex \hbox{$\sim$} \llap{\raise .27ex \hbox{$>$}} }
\def\lsim{ \lower .75ex\hbox{$\sim$} \llap{\raise .27ex \hbox{$<$}} }

\def\beq{\begin{equation}}
\def\eeq{\end{equation}}

\title{Gamma-ray burst jets: uniform or structured?}

\ShortTitle{GRB jets: uniform or structured?}

\author{\speaker{O.~S.~Salafia}$\,^{ac}$, A.~Pescalli$^{bc}$, F.~Nappo$^{bc}$, G.~Ghisellini$^c$, G.~Ghirlanda$^c$, R.~Salvaterra$^d$, G.~Tagliaferri$^c$\\
        $^{a}$Universit\`a degli Studi di Milano-Bicocca, piazza della Scienza 3, I-20126 Milano, Italy\\
        $^{b}$Universit\`a degli Studi dell'Insubria, via Valleggio, 11, I-22100 Como, Italy \\
        $^{c}$INAF - Osservatorio Astronomico di Brera Merate, via E. Bianchi 46, I–23807 Merate, Italy\\
        $^{c}$INAF - IASF Milano, via E. Bassini 15, I-20133 Milano, Italy\\
        E-mail: \email{omsharan.salafia@brera.inaf.it}}

\abstract{The structure of Gamma-Ray Burst (GRB) jets impacts on their prompt and afterglow emission properties. Insights into the still unknown structure of GRBs can be achieved by studying how different structures impact on the luminosity function (LF): i) we show that low ($10^{46} < \Liso < 10^{48}$ erg/s) and high (i.e. with $L > 10^{50}$ erg/s) luminosity GRBs can be described by a unique LF; ii) we find that a uniform jet (seen on- and off-axis) as well as a very steep structured jet (i.e. $\ept \propto \theta^{-s}$ with $s > 4$) can reproduce the current LF data; iii) taking into account the emission from the whole jet (i.e. including contributions from mildly relativistic, off-axis jet elements) we find that $\Eiso(\tv)$ (we dub this quantity ``apparent structure'') can be very different from the intrinsic structure $\ept$: in particular, a jet with a Gaussian intrinsic structure has an apparent structure which is more similar to a power law. This opens a new viewpoint on the quasi-universal structured jet hypothesis.}

\FullConference{Swift: 10 Years of Discovery\\
		 2-5 December 2014\\
		 La Sapienza University, Rome, Italy}

\begin{document}

\section{Introduction}


The idea that gamma-ray burst could be emitted by a quasi-universally structured jet (QUSJ hereafter) has been introduced more than ten years ago by Lipunov et al. \cite{lipunov-grb_standard_energy2001}, Rossi et al. \cite{Rossi-structured2002} and Zhang \& M\'esz\'aros \cite{zhang-universalconfig2002}. Very briefly, it implies that all gamma-ray burst (GRB) jets share a quasi-universal structure (\ie universal, with some dispersion of the parameters), but they appear different because we see them under different viewing angles $\tv$. Customarily, the jet structure (also called ``jet profile'') is assumed axisymmetric, and it is defined by the two functions $\ept$ (the energy emitted per jet unit solid angle) and $\Gamma(\theta)$ (the Lorentz factor of the emitting material), both depending on the angular distance $\theta$ from the jet axis.

\section{Luminosity function}

\begin{figure}
 \begin{center}
 \includegraphics[width=0.7\textwidth]{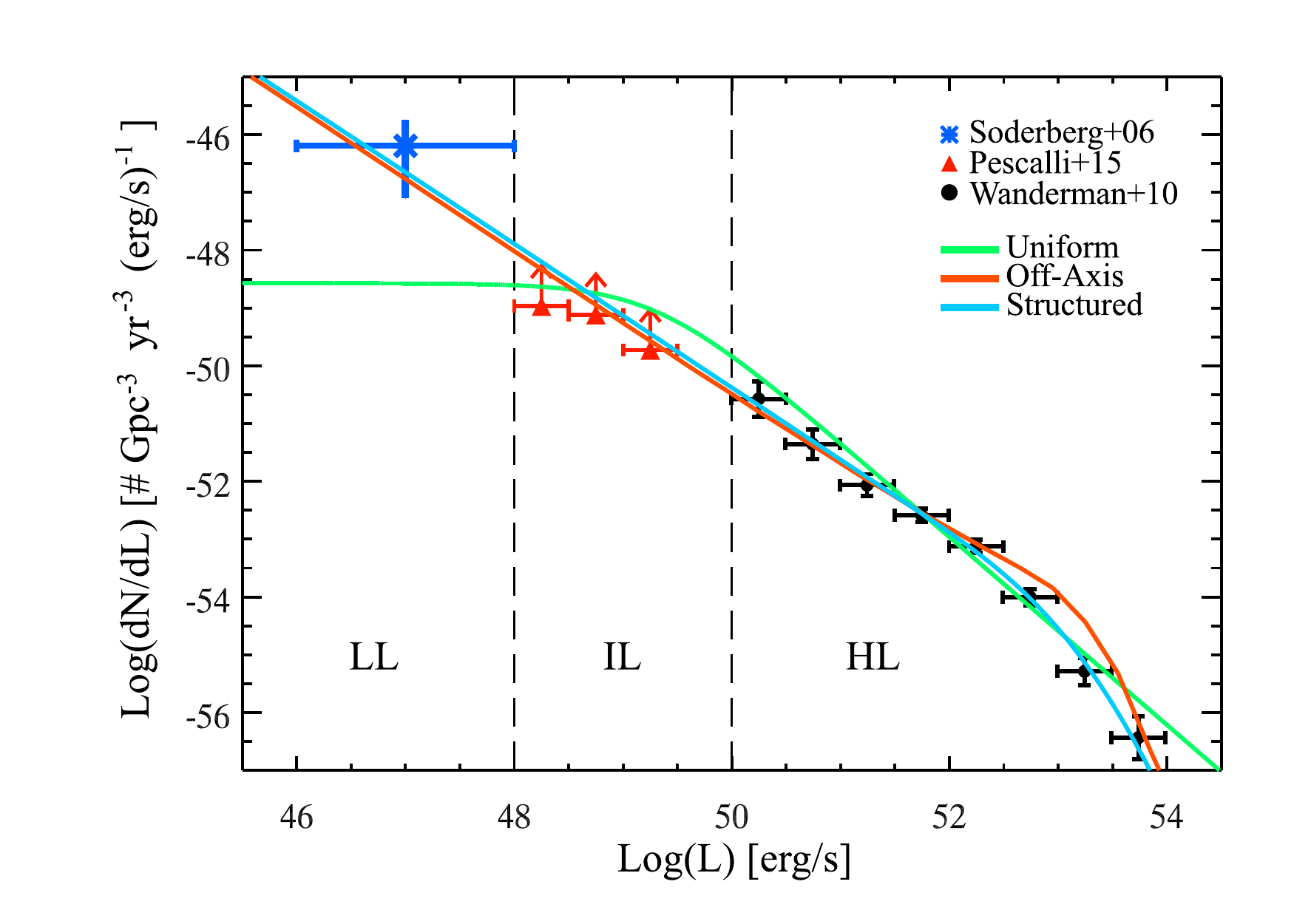} 
 \end{center}
 \caption{\label{fig:lf}Predicted luminosity functions for three jet models, along with data points from \cite{2006ApJ...638..930S},\cite{2015MNRAS.447.1911P} and \cite{2010MNRAS.406.1944W}. The green solid line is the predicted LF for a model in which the jet is uniform and observed always on--axis; the red solid line is the LF for a uniform jet that can be observed on-- and off--axis; the light blue solid line is the LF of a structured jet with $\epsilon(\theta_{\rm v}) \propto \theta_{\rm v}^{-8}$. The vertical dashed lines divide the data points in three classes: low luminosity (LL), intermediate luminosity (IL) and high luminosity (HL) bursts. }
\end{figure}

The increasing number of \GRBs with measured redshift allows to estimate their
luminosity function (LF) with increasing confidence. In a recent work \citep{2015MNRAS.447.1911P} we showed how the jet structure should affect the observed LF. Here we give a brief account of our findings.

The typical isotropic equivalent luminosities of GRBs with measured redshift are in the range $10^{50}<\Liso<10^{54}$ erg s$^{-1}$: the LF of these bursts can be modeled \cite{2010MNRAS.406.1944W} as a broken power law, with slopes $a\sim$1.2--1.5 and
$b>2$, and the break around $3\times10^{52}$ erg s$^{-1}$.

The detection rate at lower luminosities drops to only a few events: however, these events are representative of a large local density of GRBs \cite{2006ApJ...638..930S,2006Natur.442.1011P}. Some authors argued that such events could belong to a different GRB population \cite{2009MNRAS.392...91V,2007A&A...465....1D}. Conversely, in \citep{2015MNRAS.447.1911P} we showed that within some of the most common jet models the rate of low luminosity GRBs (LL GRBs) can be accounted for by the extrapolation of the luminosity function (LF) of high luminosity GRBs (HL GRBs): in this case, the complete luminosity function extends over 7 orders of magnitude.

In particular, we tested the following jet models:

\begin{enumerate}
\item first we assumed a uniform (``top-hat'') jet model, with a typical opening angle and a typical duration, always observed on--axis.
We took into account the fact that the jet has to spend part of its energy to excavate the progenitor star envelope in order to break out and produce a succesful GRB: the observed luminosity is thus proportional to the jet energy remaining after the break out. This implies that the LF (green solid line in Fig.~\ref{fig:lf}) must be flat at low luminosities, and this contradicts the data.
We thus excluded this simple case;

\item we examined then the case of a uniform jet in which the jet opening angle $\tj$ is related to the GRB energetics: the smaller $\tj$, the larger $E_{\rm iso}$, and thus $L_{\rm iso}$ (such a relation is justified by the Amati \cite{2006MNRAS.372..233A} and Ghirlanda \cite{2004ApJ...616..331G} correlations). Since lower luminosity GRB jets have larger $\tj$, the probability that they intercept our line of sight is greater than that of HL GRBs: the fraction of GRBs that we detect at lower luminosities is thus greater. Within some assumptions, we can obtain a reasonable agreement with the data in the entire luminosity range but the very low luminosities,
where the model shows a small deficit (see \citep{2015MNRAS.447.1911P} for a plot of the LF and more details on this model);

\item in the third case the jet is still uniform, but it can be observed also off--axis (\ie for viewing angles $\tv >\tj$).
Assuming some dispersion on $\tj$ and $\Gamma$, we obtained a reasonable agreement between the data and the predicted LF (red solid line in Fig.~\ref{fig:lf}).
The required average values are $\left<\Gamma\right> =30$ and $\left<\tj\right> \geq 3^\circ$;

\item finally, we investigated a QUSJ of the form 
\begin{equation}
 \epsilon(\theta) \propto \left\lbrace\begin{array}{lr}
                           \theta^{-s} &\, (\theta\geq\tc)\\
                           {\rm const.} &\, (\theta<\tc)\\
                          \end{array}\right.
\end{equation} 
We found that a good fit can be obtained, but only if the slope
$s$ is rather steep ($s>4$) with a preferred value $s\sim 8$ (light blue solid line in Fig.~\ref{fig:lf}).
This is much steeper than the value $s=2$ originally proposed by \cite{Rossi-structured2002} to explain the clustering of $E_\gamma$ found by \cite{Frail01}.

\end{enumerate}

Our results indicate that the jet must have relatively sharp edges. Even if a discontinuous structure is unphysical (all the energy uniformly distributed within $\tj$, and zero outside), the energy must in any case decrease rapidly with the angular distance from the jet axis, once this becomes greater than the core angle $\tc$. The other important overall conclusion is that while the low luminosity bursts seem not to have enough energy to punch the progenitor star, they can actually be understood within the same framework of large luminosity GRBs, as long as they have a larger jet angle (case 2 above), or as they are seen off--axis (case 3 and 4). In the former case we see the little energy leftover after the jet break--out, in the latter case the apparent low luminosity is due to the large viewing angle, but the real energetics of these burst is much greater.

\section{Apparent structure}
\label{sec:lf}

\begin{figure}
 \begin{center}
 \includegraphics[width=0.85\textwidth]{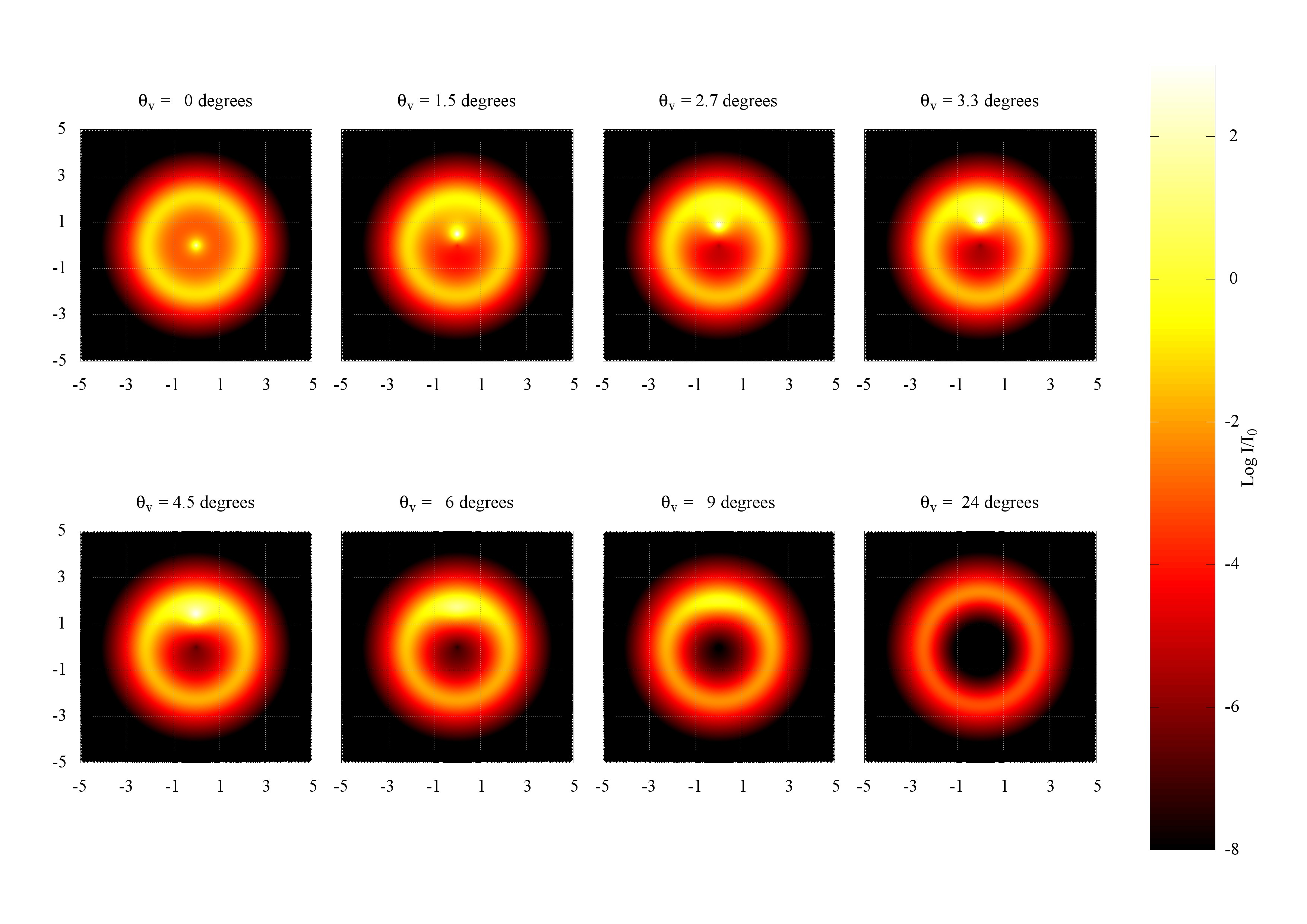} 
 \end{center}
 \caption{\label{fig:maps}Maps of the intensity emitted by each point of the jet surface are given, for a Gaussian structured jet. The relevant jet parameters are $\Gamma_c = 800$, $\tc = 3^\circ$ and $\tgtc = 1$. The color of each point represents the value of $\log \left(I/I_0\right)$ according to the colorbar on the right (values below $-8$ are rendered in black). Here $I_0$ represents the comoving intensity produced at the central point. The comoving emission is assumed isotropic. The axes of each image give the angular distance from the jet axis in units of $\tc$.}
\end{figure}

In \cite{arxiv-1502.06608} we introduced the concept of \textit{apparent structure} of a jet as opposed to its \textit{intrinsic} structure. 
While the latter is defined by the functions $\epsilon(\theta)$ and $\Gamma(\theta)$, the former is the isotropic equivalent energy $\Eiso(\tv)$ as a function of the viewing angle $\tv$. This distinction is important because many authors assume that the intrinsic and apparent structure are just proportional to each other, \ie $\Eiso(\tv) = 4\pi\epsilon(\tv)$ (\eg \cite{Rossi-structured2002,zhang-universalconfig2002,lloyd-ronning-quasi-universal-2004,2005ApJ...621..875D}), but this is not always the case. Such an assumption, indeed, is based on the fact that relativistic beaming prevents the observer from ``seeing'' those parts of the jet whose velocity forms an angle larger than $\Gamma^{-1}$ with the line of sight \cite{rybickilightman}. If the Lorentz factor $\Gamma(\theta)$ is variable, though, one can have parts of the jet that move ``slowly'' enough for their beaming cone to intercept the line of sight, even if they move away from the line of sight: in such a case, one can not neglect the contribution to the observed fluence due to these parts. Indeed, what we show in \cite{arxiv-1502.06608} is that \textit{the more the Lorentz factor varies, the less the apparent structure mimics the underlying intrinsic structure}, \ie in general $\Eiso(\tv)\neq 4\pi\epsilon(\tv)$.

The apparent structure of a jet depends on the intrinsic structure according to 
\begin{equation}
 \Eiso(\tv) = \int \dfrac{\delta^3(\theta,\phi,\tv)}{\Gamma(\theta)}\,\epsilon(\theta)\,d\Omega
 \label{eq:app-struct}
\end{equation} 
where $\delta$ is the relativistic Doppler factor. The integrand shows how to weigh the contribution from each part of the jet. In Fig.~\ref{fig:maps} we plot this integrand, for different values of $\tv$, in the case of a Gaussian structured jet defined according to \cite{kumar-sj-afterglow-2003} as
\begin{equation}
 \begin{array}{l}
  \epsilon(\theta) = \epsilon_c\;e^{-\left(\theta/\tc\right)^2}\\
  \Gamma(\theta) = 1 + (\Gamma_c-1)\;e^{-\left(\theta/\tc\right)^2}\\
 \end{array}
 \label{eq:intrinsic-structure-gaussian}
\end{equation}
Inspection of the figure shows that the bulk of the flux is not always due to the ``on--axis'' parts of the jet.

\subsection{The Gaussian QUSJ}

\begin{figure}
 \includegraphics{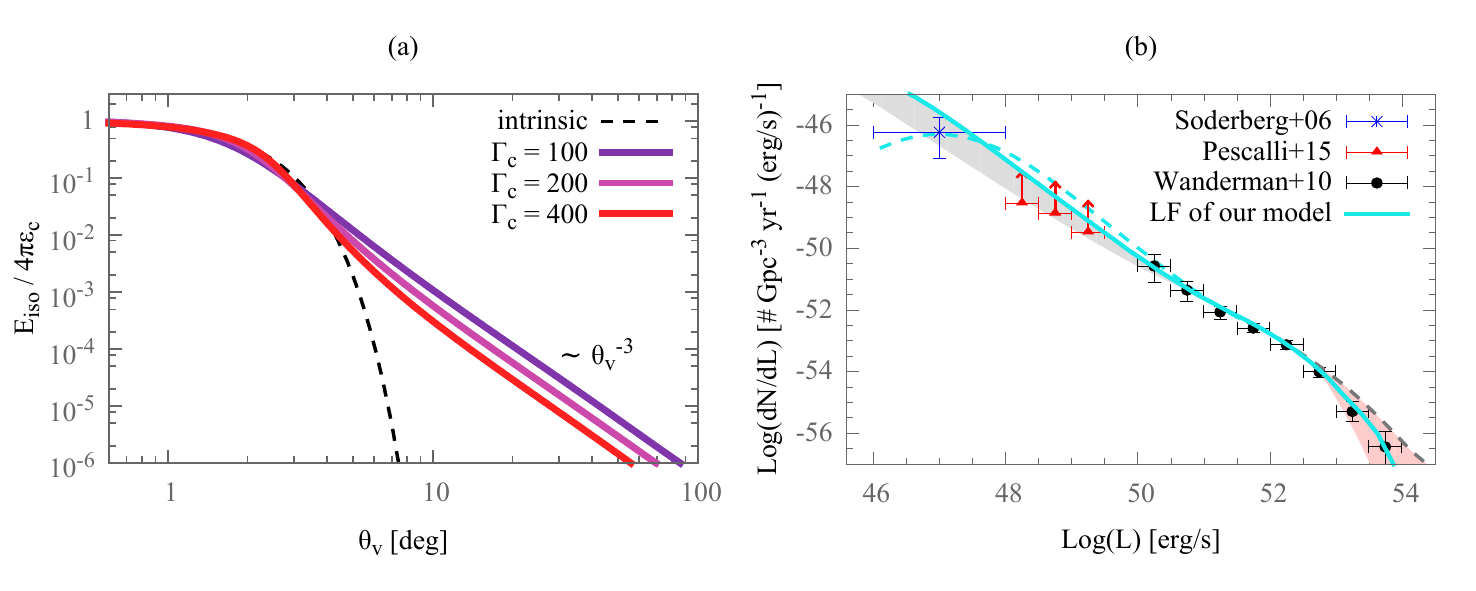}
 \caption{\label{fig:gaus-profiles}(a) apparent structures (solid colored lines) of three Gaussian jet models. The corresponding intrinsic structure (dashed black line) is drawn for comparison; (b) luminosity function of the Gaussian QUSJ model described in \cite{arxiv-1502.06608}. }
\end{figure}

The apparent structure of a Gaussian structured jet is surprisingly well described by a power law plus a roughly constant core, as shown in Fig.~\ref{fig:gaus-profiles}a. Based on this point and on the discussion in \S\ref{sec:lf}, one can argue that a Gaussian QUSJ might be compatible with the constraints posed by the observed LF, since its apparent structure, rather than the intrinsic structure, determines the LF. Indeed in \cite{arxiv-1502.06608} we show that a Gaussian QUSJ with $\tc=3^\circ$, $\Gamma_c=800$ and $\epsilon_c = 2.4\times 10^{52}$ erg/sr produces a LF (light blue solid line in Fig.~\ref{fig:gaus-profiles}b) which is in good agreement with observations, provided that the total energy of the jet has an intrinsic dispersion of about $0.5\dex$. This small dispersion points towards an unification of Long GRB jets, which is exactly what one expects by the association of Long GRBs with type Ibc supernovae, since the latter should have rather standard progenitors.

\section{Conclusions}
We showed that a steep intrinsic jet structure is a good candidate for a quasi universal \GRB jet, based mainly on its ability to predict correctly the observed luminosity function. We also showed that a clear distinction between apparent and intrinsic structure of the jets is important to predict correctly the observable properties of \GRBs.
We refer the interested reader to \citep{2015MNRAS.447.1911P} and \cite{arxiv-1502.06608} for a more complete treatment of the topics outlined above.

\bibliographystyle{JHEP}

{\footnotesize
\bibliography{journals,grb}
}

\end{document}